\documentstyle[prl,epsf,aps]{revtex}

\begin{document}
\thispagestyle{empty}

\title{Metastable states in glassy systems}

\author{Giulio Biroli}
\address{ 
\it Laboratoire de Physique Th{\'e}orique de l'Ecole Normale 
Sup{\'e}rieure\thanks{Unit{\'e} Mixte de Recherche du Centre National de la 
Recherche Scientifique et de l'Ecole Normale Sup{\'e}rieure.},
\\ 24 rue Lhomond, 75231 Paris cedex 05, France.}

\author{Jorge Kurchan}
\address{ 
\it P.M.M.H. Ecole Sup{\'e}rieure de Physique et Chimie Industrielles,
\\
10, rue Vauquelin, 75231 Paris CEDEX 05,  France}

\date\today

\maketitle

\begin{abstract}
 Truly stable metastable states are an artifact of the mean-field
approximation or the zero temperature limit. If such appealing
concepts in glass theory as configurational entropy
are to have a meaning beyond these approximations, one needs to
cast them in a form involving states with finite  lifetimes.
 Starting from elementary examples and using results of Gaveau
and Schulman, we propose a simple expression for the 
configurational entropy and revisit the question of 
taking flat averages over metastable states.
The construction is applicable to
finite dimensional systems, and we explicitly show that
 for simple mean-field glass models
  it recovers, justifies and generalises the known  results.
  The calculation emphasises the appearance of new dynamical
 order parameters.

\end{abstract}
\vspace{.5cm}
 

 \section{Introduction}

Slowly relaxing systems such as  glasses or  compacting granular media
can be viewed as having  fast, local, quasi-equilibrium dynamics,
plus a slow, nonequilibrium drift.
These two superposed motions can take different forms: `cage'
vibrations plus structural rearrangements in glasses;
bulk fluctuations plus domain-wall motion in coarsening problems; etc.
At a given long time, the fast motion covers a region of phase space
which one may picture as  a `metastable state'.

Even though metastable states are a familiar and appealing concept,
it turns out that defining them in an  unambiguous manner in non-mean
field
models is quite
subtle. This in turn has as a consequence that such
standard ideas in glass
theory as `configurational  entropy' (related to the number of 
metastable states) are not only hard to calculate, but are
indeed, with the exception of some fortunate cases, approximate
  {\em as concepts}.

 In this paper we show how these questions can be put on a well
 defined basis using a formalism \cite{larry} that does not rely on
 specifically  mean-field concepts.

First of all: given that one can simulate and in certain cases
calculate analytically the complete history of a sample starting from
a quench, why should one have any need to introduce the apparently
unneeded notion of `metastable state' ?
Indeed, these  states 
only come into play when one wishes to make arguments such
as: {\em  `phase space contains such and such a distribution of states,
which will be accessed with such and such a probability by a typical
dynamical history. Long-time out of equilibrium
observables can be directly
calculated by averaging the observables
 over some subset of states --- and further
reference to dynamics may be  omitted'}.

This kind of `ergodic' argument was
pioneered by Edwards \cite{Sam}, who proposed that in compacting or
slowly flowing granular systems one can obtain the correct 
dynamical observables by averaging the values they take over all blocked 
configurations  of a certain volume.
It later turned out that mean-field glass models \cite{Beyond,review}
relaxing at 
zero temperature had exactly Edwards' ergodicity property
\cite{jorge-trieste}: at long
times any nonequilibrium observable is correctly given
by the typical value it takes over all local energy minima of the
appropriate energy density.

A first  problem arises when one wishes to apply this concept at finite 
temperatures (or vibration, in the case of granular media).
 There again, the mean-field case offers a suggestion:
 at non zero temperature Edwards' argument works as well, provided one
substitutes `energy minima' by `free-energy minima' (`states'). 
This construction is possible because within mean-field we have a 
well defined notion of free-energy landscape, whose local minima
are in some (but not all) cases related to completely stable distributions.
However, as discussed by Franz and Virasoro \cite{Frvi}, one
 needs to consider `quasi-states' with finite lifetimes
in order to understand the situation at finite waiting times.

In finite-dimensional problems, a high-lying metastable state
cannot have an infinite lifetime: there is always a finite probability 
of escape through the nucleation of  droplet of a more favourable
phase. 
 Hence, which distribution one considers as metastable depends always
on which lifetimes one is considering.
For example, the concept of  `configurational entropy' (the logarithm
of
the number of states), ubiquitous in glass theory, has in finite
dimensions only a meaning with a timescale attached.
 Moreover, even the mere definition of an Edwards distribution 
is not as simple,  quite apart from the question of the 
validity of the  ergodicity-like hypothesis it assumes.

In section 2 we shall review the notion of metastable state within
mean-field glass models and how the knowledge
of their distribution allows in certain cases to reproduce some  results 
obtained from the full solution of the out of equilibrium dynamics.
We shall also mention some limitations found even at this level
 of the identification `free-energy minimum $\sim$ stable state'.

In section 3 we discuss a strategy valid in any dimension
for the definition and calculation
of metastable states based on the evolution operator developed
by Gaveau and Schulman \cite{larry}.
We discuss how one can thus  recast the question of configurational
entropy
 and Edwards' distribution in a form relevant in finite dimensions
at nonzero temperature (or stronger vibration, in the case of
granular media) by considering finite-lifetime metastable states,
in the spirit of the `quasi-states' discussed in \cite{Frvi}.

In section 4 we apply this method to a simple mean-field  glass model.
We  show how one can rederive  in this way both the number of states 
and the dynamics inside a state,  
within a framework whose applicability goes beyond  mean-field.
 
\vspace{.5cm}

\section{A Fortunate case: mean-field models}

\vspace{.5cm}

Consider the mean-field model of ferromagnet:
\begin{equation}
E=-\frac{1}{2N} \sum_{i,j} S_i S_j -h \sum_{i} S_i
\label{curie}
\end{equation}
where the sum is over all spins. The spins can be Ising
  $S_i =\pm 1$ or spherical $\sum_i S_i^2=N$.
One can easily obtain a free energy in terms of local magnetisations
$m_{i}=\left< S_i\right>$ (where $\left<\cdot \right>$ means the average over
 the Gibbs measure):
\begin{equation}
f(\{m_{i} \})=-\frac{1}{2N}\sum_{i\neq j} m_i m _j -h \sum_{i} m_i - T S(\{m_{i} \})
\label{curie1}
\end{equation}
where  $T=1/\beta $ denotes the temperature and 
$S(\{m_{i} \})$ is  the usual entropic term: 
\begin{eqnarray}
&S(\{m_{i} \})=&- \sum_{i}\frac{1}{2} \left[ (1-m_{i})\ln(1-m_{i})+(1+m_{i})\ln(1+m_{i})
 \right]
\;\;\;\; {\mbox{for}} \;\;\; S_i=\pm 1 \nonumber \\
&=& \frac{1}{2} \ln (1-q)
 \;\;\;\;\;\;\;\;\;\;\;\;\;\;\;\;\;\;\;\;\;\;\;\;
\;\;\;\;\;\;\;\;\;\;\;\;\;\;\;\;\;\;\;\;\;\;\;\; {\mbox{ spherical model}}
\label{entr}
\end{eqnarray}
and $Nq=\sum_{i=1}^{N}m_{i}^2$.

The states are represented by the minima of $f$. At $T<T_c$ there
are two, and the deeper
one dominates the Boltzmann distribution.
If $h>0$ one of the states becomes metastable: within mean-field 
its lifetime is infinite, but in finite dimensions it will decay.


\vspace{.5cm}

{\it The TAP approach}

\vspace{.5cm}

Next, consider the glassy Hamiltonians:
\begin{equation}
E=-\frac{1}{p!} \sum_{i_1,...,i_p} J_{i_1,...,i_p} S_{i_1}... S_{i_p}
\label{pspin}
\end{equation}
where the $J_{i_1,...,i_p}$ are independent random variables with variance
$p!/2N^{p-1}$.
 The Ising version with $p=2$ is the Sherrington-Kirkpatrick model,
the mean field version of {\em spin}-glasses. The models with 
$p \geq 3$ are instead systems having the behaviour resembling
{\em structural} glasses \cite{KTW,MP}. 

 It turns out that one  can find for these models a free energy
 function analogous to (\ref{curie1})
(the `TAP' free energy \cite{TAP}),
  in terms of local magnetisations $m_i$:
\begin{equation}
f_{TAP}(\{m_{i} \})= -\frac{1}{p!} \sum_{i_1,...,i_p} J_{i_1,...,i_p}
 m_{i_1}... m_{i_p}- \frac{\beta}{4} [1-q^p-pq^{p-1}(1-q)]- T S(\{m_{i} \})
\label{TAP}
\end{equation}
where $Nq=\sum_i m_i^2$ and $S(\{m_{i} \})$ is given in (\ref{entr}) for
the spherical and Ising cases, respectively.

Within the TAP approach, one signature of the glass transition 
is the fact that the  free energy (\ref{TAP}) has, below a critical
temperature, many ($ \sim e^{aN}$) minima\cite{Dedominicis,BM,crisomtap,Cagipa}.
The main difference between mean-field versions of {\em spin} and 
{\em structural} glasses is seen, in the TAP approach, in the way states 
are separated.

Given the analogies with the ferromagnet, it seems very tempting
to attribute to the minima of (\ref{TAP}) a dynamical
meaning of  `state'. 
For the models with $p \geq 3$ this has been done \cite{Bamebu,tapdyn}
by starting from a configuration where the coordinates are as close as
possible to having a given $m^a_i$, 
and then checking that the subsequent dynamical
evolution is a stable, quasi-equilibrium situation confined to a
 region of phase space in such a way that  
$<S_i>_{\mbox{\small{time-average}}}=m^a_i$.

On the contrary, 
for models like the Sherrington-Kirkpatrick model the identification of 
all TAP-minima with  stable states  seems to breaks down. 
Let us formulate an heuristic argument to see this. Decreasing  temperature,
minima split in a second-order transition manner.
At least a fraction of the minima  are `born' this way \cite{Dotsenko},
and to get an exponential number of minima one needs that on average 
there is a division every $O(1/N)$ change in temperature. Hence, a fraction
of TAP solutions are just $O(1/N)$ below their critical temperature,
and under those circumstances  barriers cannot be large enough to
dynamically separate them from their `twins'. 

\vspace{.5cm}

{\it FDT-temperature. Long-time out of equilibrium observables}

\vspace{.5cm}

The dynamics of  model (\ref{pspin}) following a quench  below the
critical temperature can be solved analytically \cite{Cuku,review}.
One finds that the system never equilibrates, and remains aging
just above a threshold level of energy density $e_{th}$ and free energy 
density $f_{th}$ higher than  
the equilibrium ones.
Given any two observables $A$ and $B$, one can define their dynamic
correlation function:
\begin{equation}
C_{AB}(t,t')= \langle A(t) B(t') \rangle
\label{Ceq}
\end{equation}
and the integrated response $\chi_{AB}$ to a field $h_B$ conjugate to $B$
acting between times $t'$ and $t$:
\begin{equation}
\chi_{AB}(t,t')= \frac{\delta \langle A(t)  \rangle}{\delta h_B}
\label{Req}
\end{equation}
where the averages are over the dynamical realisation.

\begin{figure*}
\vspace{.3cm}
\centerline{\input{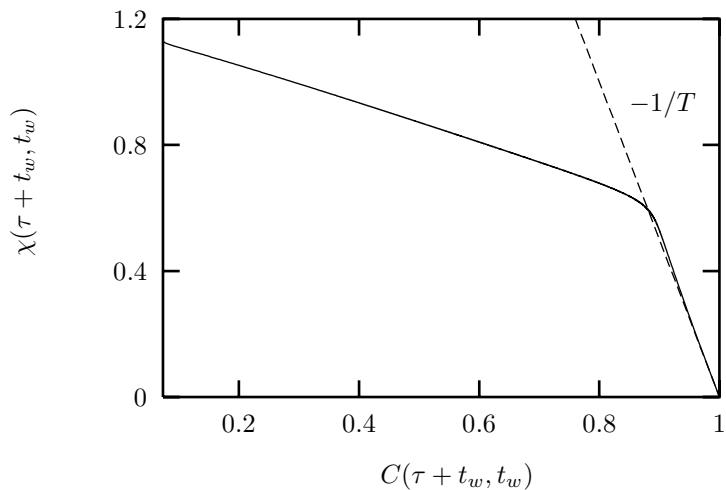}}
\vspace{.3cm}
\caption{A fluctuation-dissipation plot. 
The straight line to the left defines the effective
temperature.}
\label{fig1}
\end{figure*}

If one makes a parametric plot of $\chi_{AB}(t,t')$  versus  $C_{AB}(t,t')$
one obtains for model (\ref{pspin}) with $p>2$ at long times a curve like Fig. 2.
One has in addition to the  straight line with gradient $-1/T$
(as in equilibrium), another straight line of gradient, say,
$-1/T_{eff}$,
associated to the slow relaxation. 
 $T_{eff}$ is the same for every pair of observables $A$ and $B$, and
 it can be shown \cite{Cukupe} that it satisfies all the properties
of a true temperature.

The appearance of a temperature in a system is an indication of some
form of ergodicity, in this case clearly not the usual Gibbs-Boltzmann
equilibrium at temperature $T$.
Indeed, soon after the dynamical solution was obtained, Monasson and
Virasoro \cite{Remi,Virasorounpu} 
observed that the temperature $T_{eff}$ could be reobtained
from the TAP approach without making any reference to the dynamics:
 Defining the complexity (or configurational entropy) ${\cal{S}}(f)$
as the logarithm 
of the number of TAP solutions of a given free energy, one checks that
\begin{equation}
\frac{1}{T_{eff}}= \left. \frac{\partial  {\cal{S}}(f)}{\partial f}
\right|_{f=f_{th}}
\label{gr}
\end{equation}
How this equality follows from Edwards' assumption is discussed in 
\cite{Frvi}.

 Furthermore, one can also see that the long-time values of
 macroscopic observables are given by the {\em flat}
 average of their values taken over all TAP solutions of the
 dynamical energy density $e_{th}$.
Hence, we have a strong indication for a  measure {\`a} la Edwards,
this time applied to TAP states.

If the system has large, but finite size, it will slowly approach
equilibrium.
In that case, a plot like Fig. 2 will show that the two  tracts
slowly
tend to become parallel and $T_{eff}(t)$ (now a function of time)
tends to $T$.
Inspired by work of Bonilla et. al. \cite{Bopari}, Nieuwenhuzen \cite{Theo}
conjectured that one could extend the relation (\ref{gr}) for all
times, using the TAP solutions of the energy level appropriate at each time.

Later on, a two-temperature $\chi$ versus $C$ plot (and hence the
 existence of an FDT temperature $T_{eff}$) was seen to occur
 in realistic models such as three dimensional Lennard-Jones glasses 
\cite{Pa,Kobteff,aggiunta2,Ru}. 
Several of these simulations where performed at temperatures {\em above} the
 putative equilibrium glass temperature, so that the existence of a
(slowly evolving) well-defined $T_{eff}$  is surely 
a dynamical phenomenon,  unrelated to the structure of
 equilibrium states.
If one wishes to consider this as a symptom of slowly evolving
flat distribution 
between metastable states,  one finds oneself in the embarrassing situation that
it is not entirely obvious what one means by `metastable 
state' in finite dimensions, as we now discuss.

\vspace{.5cm}

{\it The problem with finite dimensional and driven systems}

\vspace{.5cm}

In finite dimensions and non-zero temperature 
nucleation arguments suggest that 
a distribution with dynamical free energy density (to be defined
below) higher than the equilibrium one 
should decay through nucleation in finite times.  
We are hence in a situation in which we have no absolute notion of
state without making reference to a timescale (and hence to  dynamics):
 two different distributions may be confused into a single state
or be treated as two separate entities  depending on whether
 the time to go from one to the other is smaller than or larger than
the  timescale considered.

If we are interested in systems driven by shear or by vibration, 
we have the additional problem that even in the mean-field case
the distribution is not Gibbsean within a state.  
In a vibrated case, the notion of stability must be substituted by the
notion of periodicity, so that a `state' will turn out to be a
structure periodic in time. 

Before entering into the present approach, let us mention that a
pragmatic way of dealing with these difficulties, at least at 
 very short time-scales, is the so-called
`inherent structure' construction \cite{inherent}. 
Though it does not solve the questions of principle mentioned above 
\cite{enfants}, 
it offers a practical way around applicable  to concrete problems.

\vspace{.5cm}

\section{Dynamical definition of metastable states}

\vspace{.5cm}

Let us consider a system evolving with stochastic dynamics, which for
definiteness we shall consider is of the  Langevin form.
The probability distribution will evolve according to:
\begin{eqnarray}
\frac{dP(S,t)}{dt}&=& -H P(S,t) \nonumber\\
H &=&-\frac{\partial}{\partial S_i} \left(T\frac{\partial}{\partial S_i}+
\frac{\partial E}{\partial S_i}\right)
\label{fokker}
\end{eqnarray} 
where $H$ is the Fokker-Planck operator.
The potential energy $E$ can be time-dependent, and furthermore
one can add to $(\frac{\partial E}{\partial S_i})$  forces
that do not derive from a potential.

Given a probability distribution $P$, one can define
a dynamic free energy
\begin{equation}
F(t)= \int \left( T P(S,t) \ln P(S,t) +E(S)P(S,t) \right)  \;dS
\end{equation}

If $H$ is time-independent, any  stationary configuration 
satisfies
\begin{equation}
H P_{\mbox{\small stationary}}=0
\end{equation}
Moreover, writing any distribution as
$P(x,t)= \sum c_i(t) \psi_i(x)$ where $\psi_i(x)$ are the right
eigenvectors of $H$:
\begin{equation}
H |\psi_i\rangle  =\lambda_i  |\psi_i\rangle
\end{equation}
the evolution equation (\ref{fokker}) implies:
\begin{equation}
c_i(t) =  c_i^o e^{-\lambda_i t}
\label{sol}  
\end{equation}
We see that if $P$ is to vary slowly it has to be concentrated on
eigenvectors with low eigenvalues $\lambda_i$. Indeed, each
 $\lambda_i$ is an inverse timescale \cite{regis,spectre}.

In the following subsections we shall motivate and discuss
an identification of the set of small eigenvalue $\psi_i$'s 
with metastable states.

\vspace{.5cm}

{\it Motivation}

\vspace{.5cm}

\begin{figure}
\centerline{\epsfxsize=9cm
\epsffile{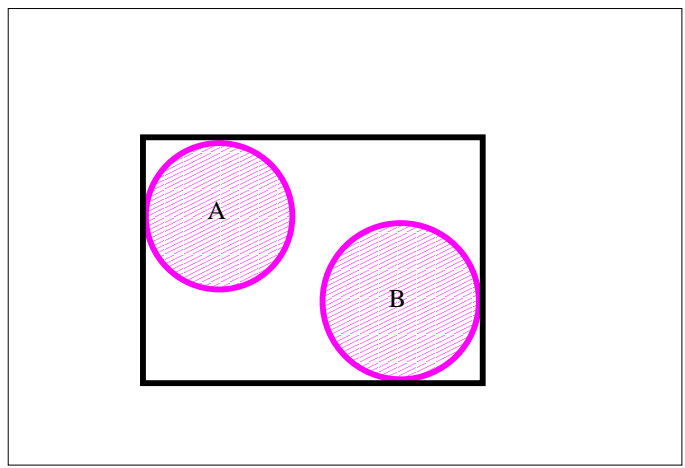}
}
\caption{A system with two ergodic components.}
\label{coins}
\end{figure}

Consider first the system of two hard disks performing Langevin 
dynamics in a box (Fig. \ref{coins}).  Clearly, if the disks are really 
not interpenetrable, there are two different ergodic components, each
composed of the mirror image of the configurations of the other.
Symmetry implies that 
the spectrum of the Fokker-Planck operator is
doubly degenerate. The two lowest (zero) eigenvalues correspond to two
stationary distributions. One can
construct an associated  eigenvector as the flat distribution over all
pairs of coordinates of the centers of the disks such that they do not
superpose and such that 
the disk `A' is to the right and `B' to the left, and similarly a second
eigenvector corresponding to having the disk `B'  to the right and `A' to the left.
These are the `pure states': any linear combination of these two
distributions will be an `impure' state.
The next higher eigenvalues are equal to the inverse of the time
needed for the particles to explore their ergodic component. 
Note, in passing, that this hard-spheres system is the typical example
in which the inherent structure construction is not meaningful 
while the present one has no problem.

In this example we have strictly two ergodic components,
corresponding to twofold  ground state degeneracy of the Fokker-Planck
operator. This is indeed very general: suppose we have $p$ ergodic
components ${\cal{C}}_1,...,{\cal{C}}_p$,
 with typical times $t_1,...,t_p$ required to explore
  each component. We can construct an independent eigenvector with
zero eigenvalue using the stationary distribution $P_a(x)$ restricted to each
component ${\cal{C}}_a$.
These completely span the zero eigenvalue subspace. To show this, we calculate:
\begin{eqnarray}
{\mbox{tr}} [e^{-t^*H}] &=& \int \; dx \int \; dy \langle x
|e^{-\frac{t^*}{2}H}|y \rangle \langle y
|e^{-\frac{t^*}{2}H}|x \rangle =
\sum_{a=1}^p \sum_{b=1}^p \int_{x \in {\cal{C}}_a} dx 
\int_{y \in {\cal{C}}_b} dy \;  \langle x
|e^{-\frac{t^*}{2}H}|y \rangle \langle y
|e^{-\frac{t^*}{2}H}|x \rangle \nonumber\\ &=&
\sum_{a=1}^p \int_{x \in {\cal{C}}_a} dx \int_{y \in
  {\cal{C}}_a} dy \;  \langle x
|e^{-\frac{t^*}{2}H}|y \rangle \langle y
|e^{-\frac{t^*}{2}H}|x \rangle
\end{eqnarray}
 If we now take  $t^*$
much larger than all the $t_i$,  $\langle y
|e^{-\frac{t^*}{2}H}|x \rangle\simeq P_a(y)$, the  equilibrium
probability for $y$ restricted to the ergodic component $a$ to which $x$ belongs.
Hence:
\begin{equation}
{\mbox{tr}} [e^{-t^*H}] \simeq  \sum_{a=1}^p
\sum_{b=1}^p \int_{x \in {\cal{C}}_a} dx 
\int_{y \in {\cal{C}}_a} dy \; P(x) P(y) \simeq p
\end{equation}
This shows that the number of states `below the gap' coincides with the number of ergodic
components.

\begin{figure}
\centerline{\epsfxsize=9cm
\epsffile{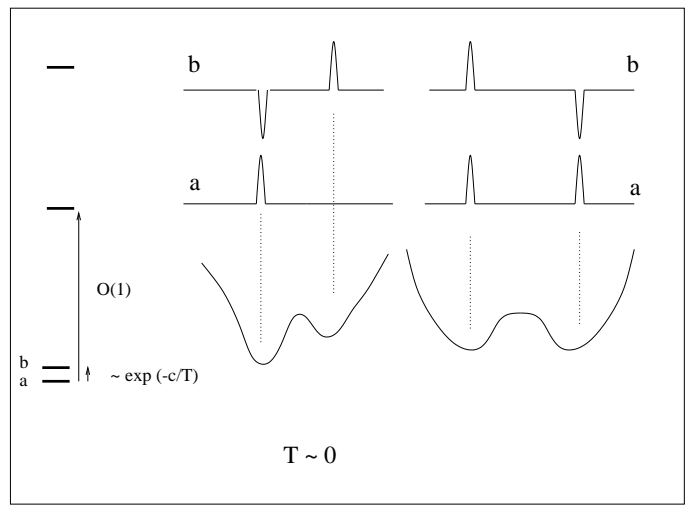}
}
\caption{A sketch of eigenfunctions and spectrum of the
  Fokker-Planck
operator at low temperature .}
\label{figg}
\end{figure}

In the preceding examples the ergodic components are strictly
separated. However, in most applications this is not the case: there
is
in fact a passage time between components that only becomes infinite in
some limit.
To understand the construction in these cases,
consider a very low temperature Langevin process 
occurring in asymmetric and symmetric double-well potentials
as in Figure \ref{figg}. 
On the left of the figure we show the lowest levels of the spectrum 
for both cases, and at the top the corresponding eigenvectors.
For the asymmetric case, the  two lowest eigenvalues are separated 
by the inverse Arrhenius escape time from the highest minimum. 
All other eigenvalues are much higher ($O(1)$), and include
the escape time from a maximum, etc. 
The 
eigenfunction labeled {\bf a}  is   essentially
positive and represents a ``pure'' state $P_1$, while one can make a linear
combination $P_2=({\mbox {\bf a}}+ {\mbox {\bf  b}})$
 that will also be positive and concentrated on the metastable minimum. 
For the symmetric well the situation is similar, but now it
is the linear combinations  $P_1=({\mbox {\bf a}}+ {\mbox {\bf  b}})$
and $P_2=({\mbox {\bf a}}-{\mbox {\bf  b}})$ that play
the role of ``pure'' states.
Any other combination of the form 
$y P_1 + (1-y) P_2$ with $0<y<1$ will give
an `impure' (almost) steady state.

Note that these definitions make sense in the time-window in which
we can consider the exponential Arrhenius times much larger that 
any other time involved ($O(1)$), and this will happen only
in the low temperature limit.
 
If the temperature is non-zero, a separation of timescales can happen
as a result of the thermodynamical (or other) limit.
For example, it is easy to see that a mean-field ferromagnet
at $0<T<T_c$ will have a similar spectrum, but with $N$ playing
the role of large parameter instead of $1/T$.
The case of finite dimensional ferromagnets is slightly more subtle:
we have there a timescale for domain excitations within a state
that can be as large  as a power law in $N$ (the time it takes for a
large domain to collapse), and a much longer 
timescale ($ \sim e^{c N^{(d-1)/d}}$) for going from one phase to the other.

\vspace{.5cm}

{\it The construction of Gaveau and Schulman}

\vspace{.5cm}

In general, the low eigenvalue spectrum can correspond to
more than two eigenvectors. 
One can now ask in general whether all these eigenvectors (or
combinations of them)
represent positive, stable  distributions, and whether one can
construct as many pure states as there are low eigenvalues.
In a series of papers\cite{larry}, Gaveau, Schulman and Lesne
showed this to be the case,  gave recipes  for the explicit 
construction of the pure states,  proved their unicity
 and exploited  this construction to study metastability.

We shall be only descriptive here, we refer the reader to the
 references for the proofs, as well
as other investigations concerning metastability.
Consider for example a Fokker-Planck operator having
the lowest $p$ eigenvalues $\lambda_1,...,\lambda_p$
separated by a gap from the others, {\i.e.}
one can find a $t^*$ such that one can consider that:
\begin{eqnarray}  
t^*\lambda_i &<<1& \;\;\;\; {\mbox{for}}\;\;\; i=1,...,p \nonumber\\
t^*\lambda_i &>>1& \;\;\;\; {\mbox{for}}\;\;\; i=p+1,...
\label{timescale}
\end{eqnarray}
In the previous simple example any $t^*$ such that  $t^*$ is positive will
satisfy this as $T \rightarrow 0$.
The meaning of $t^*$ is clear: it is a timescale
much longer than the  relaxation into states, but much shorter than
transitions between states.
Clearly, the operator $\exp[-t^*H]$ is essentially a projector onto
the space `below the gap' (up to terms of order $\exp[-t^*\lambda_i]$,
with $i\leq p$). 
Within the same accuracy, one can then find a basis of $p$ right 
eigenvectors $|P_i\rangle$ which are:
\begin{itemize}
\item positive:
  $P_i(x) = \langle x |P_i\rangle\geq 0$:
\item almost stationary: $H |P_i\rangle \sim 0 \;\;\;\;\;\; \forall \;
  i=1,...,p $
\item normalised and not zero in non-overlapping regions of space.
\end{itemize}
The last property is related to the fact that one can
also find a basis of $p$ approximate almost stationary
($\langle  Q_i| H \sim 0$) left eigenvectors $ \langle  Q_i|$, 
such that each $Q_i$ is essentially one within the support of $P_i
(x)$, zero everywhere else and satisfy the orthogonality and
normalisation conditions:
\begin{equation}  
 \langle  Q_i|P_j\rangle \sim \delta_{ij} 
\label{a}
\end{equation}
Given  any observable $A$, we can calculate its average within the
state ``$i$'' as:
\begin{equation}  
\langle A \rangle_i = \langle  Q_i|A|P_i\rangle 
\label{expectation}
\end{equation}
One can also write  approximately:
\begin{equation}  
 e^{-t^* H} \sim \sum_i |P_i\rangle\langle  Q_i|
\label{generator} 
\end{equation} 
As a consequence the $|P_i\rangle$ vectors have all the good properties
 to represents metastable states: they are positive normalised distributions, 
 non zero only on different regions of the configuration space and they
 are stationary on time scales less than $t^{*}$. 
In the proof, as in the simple example of the previous subsection,
 the definition
is unavoidably linked to a timescale: if one considers really
infinite times, before any other limit, then the distinction between
states vanishes.
Furthermore, it is
 assumed the number of states so defined
remains finite in the thermodynamic limit.
We believe that going to situations  in which this is not the case (as we
will below) is indeed not entirely innocent, but is at the heart of
quite a few problems associated to the definition of ``state''
 in glassy systems. 
(We have already encountered such subtleties when we discussed TAP
minima in the Sherrington-Kirkpatrick model).
\vspace{.5cm}

{\it Driven systems.}

\vspace{.5cm}

The construction described above is not limited to purely relaxational
or time-independent systems.
Consider for example the case in which a system is periodically
`vibrated' or `tapped'. One can still try to look for stationary, 
or rather, periodic situations. 
One can repeat essentially the same argument
by considering the evolution operator through one cycle:
\begin{equation}
U= {\cal{T}} \int_{\mbox{\small cycle}} dt \; e^{-t H(t)}
\end{equation}
where ${\cal{T}}$ denotes time-order.
One has to look now for the eigenvectors of $U$ whose eigenvalue is
close {\em to one}.

Similarly, one can also work with systems driven by constant
nonconservative forces (as in a sheared fluid), and with nonlinear
space-dependent friction (as in granular systems).

\vspace{.5cm}

{\it Pure barriers.}

\vspace{.5cm}

Given that  pure states can be viewed as playing the role
for  finite temperature that certain energy minima play for zero temperature,
one is naturally led to ask which distributions 
play the role of barriers (or in general saddle points) in 
finite temperature. 
For Fokker-Planck processes this can be done 
naturally starting
from the states constructed as above. In Appendix A we show how this
can be done.

We remark that a solvable example, in which saddle points play an 
important role
 in the spectrum of the evolution operators, is the Glauber 
evolution of the completely connected
 Ising model and its generalizations \cite{spectre}. It has been found that
 the lowest eigenvalues, i.e. the longest relaxation times, are gathered  
 in families, each one being in correspondence with a stationary, {\it not
 necessary stable}, point of the static mean field free energy.
\vspace{.5cm}

\section{Flat  distribution over {\em states} at finite temperature or
  vibration. A definition of configurational entropy.}

\vspace{.5cm}

\vspace{.5cm}

{\it Motivation.}

\vspace{.5cm}

As we mentioned in Section II, the fact that one finds a
 two-temperature
behaviour in mean-field glasses can be seen as suggesting the
 relevance of   
a  measure consisting in summing over  metastable states
of given energy (or free energy) with equal weight. 
Armed with the construction for metastable states we discussed above,
 we shall
see how this measure can be expressed in finite dimensional, or
 periodically driven 
systems.

\vspace{.5cm}

{\it Expressions.}

\vspace{.5cm}

Consider, in the spirit of Edwards' distribution, an
average of an observable over states, each measured with equal weight.
This will be relevant for the long-time out of equilibrium dynamics
under the assumption that almost all states of given
energy have the same basin of attraction.

Using the same notation as in the previous section, we define the
average over the measure $\langle \langle\cdot \rangle\rangle $ of an observable A 
in the following way:
\begin{equation}
\langle\langle A\rangle\rangle_{t^*} = \frac{1}{p} \sum_{i=1}^p \langle A \rangle_i
\label{edwards}
\end{equation}
where the subindex $t^*$ reminds us of the fact that states are now
defined according to their timescale. 
We have:
\begin{equation}  
\sum_{i=1}^p \langle A \rangle_i = \sum_{i=1}^p\langle  Q_i|A|P_i\rangle 
=\sum_{i=1}^p {\mbox{tr}} \left[|P_i\rangle \langle  Q_i| A\right]= {\mbox{tr}} [e^{-t^*H} A]
\end{equation}
where we have used (\ref{expectation}) and (\ref{generator}).
Hence:
\begin{equation}
\langle\langle A\rangle\rangle_{t^*} =
 \frac{{\mbox{tr}} [e^{-t^*H} A]}{{\mbox{tr}} [e^{-t^*H}]}
\label{edwards1}
\end{equation}
Note that once written this way, all reference to pure states has
disappeared, except indirectly in the value chosen for $t^*$.

We often need an equation like  (\ref{edwards1}), but restricted 
to states having a certain energy, particle number, etc.
In that case we generalise   (\ref{edwards1}) as, for example:
\begin{equation}
\langle\langle A\rangle\rangle _{t^{*}}(E_o) = 
\frac{{\mbox{tr}} [\delta (E-E_o)\;e^{-t^*H} A]}{{\mbox{tr}} [\delta (E-E_o)\;e^{-t^*H}]}=
\frac{{\mbox{tr}}_{E_o} [e^{-t^*H} A]}{{\mbox{tr}}_{E_o} [e^{-t^*H}]} 
\label{edwards2}
\end{equation}
where ${\mbox{tr}}_{E_o}$ denotes a restricted  trace.

Once we make the assumption that all states of the same energy
(or particle number, etc) have an equal weight for
the purposes of calculating a dynamical observable, it
becomes meaningful to count their number  at given energy,
the configurational entropy:

\begin{equation}
{\cal {S}}_{t^*}(E_o) \equiv \ln {\mbox{tr}} [\delta (E-E_o)\;e^{-t^*H} ]=
\ln {\mbox{tr}}_{E_o} [e^{-t^*H}]
\label{compl}
\end{equation}
If  $t^* \rightarrow 0$ we get the microcanonical measure (and entropy, up to irrelevant
constants \cite{foot}), and  if we
let $t^* \rightarrow \infty$ we find no   high-lying
metastable states at all in finite dimensions.
The dependence on $t^*$ is hence unavoidable if one is to obtain a 
finite  configurational entropy in that case.

Equation (\ref{compl}) defines the timescale-dependent configurational entropy.
One also needs the average entropy 
{\em within a state} ${s_{t^*}}$,  and the corresponding
average free-energy of a state $f_{t^*}=E_o-T{s_{t^*}}$. Using  the construction of 
(\ref{a}), (\ref{expectation})
and (\ref{generator}), we have:
\begin{equation}
 s_{t^*}(E_o)= -\frac{\int_{x/E(x)=E_o}  dx \; \langle x|e^{-t^*H}|x\rangle \; 
\ln [\langle x|e^{-t^*H}|x\rangle] } {\int_{x'/E(x')=E_o}  dx' \; \langle x'|e^{-t^*H}|x'\rangle }
\label{within}
\end{equation}
The meaning of this equation becomes transparent in the example of the completely separated
ergodic components of the previous section.

{\em Note that also the intra-state entropy is timescale-dependent}. Indeed, if we set
 $t^* \rightarrow 0$,
we are defining as `states' the configurations themselves \cite{foot}.
 On the other extreme, if $t^*$ is longer than the equilibration time, 
$\langle x'|e^{-t^*H}|x'\rangle$ gives the Gibbs-measure, and the 
intra-state entropy
becomes the usual entropy.
In short: changing the timescale both changes the number of states
and their nature, hence the change in 
configurational and intra-state entropy, respectively.

\vspace{.5cm}

{\em Flat measures and effective temperatures}

\vspace{.5cm}

Suppose we have a glassy system, taken through a given thermal history
(an annealing protocol)  to a glassy phase at time $t$, at which time
its energy 
  is $E(t)=E_o$ (and if we let the
particle number or the volume change we should specify also them). 
The assumption of typicality of metastable states, is then:
\begin{equation}
\langle\langle A\rangle\rangle_{t^*}(E_o) =
\frac{{\mbox{tr}}_{E_o} [e^{-t^*H} A]}{{\mbox{tr}}_{E_o} [e^{-t^*H}]} 
\sim \left< A \right>_{{\mbox{\small history}}}
\label{edwards4}
\end{equation}
where the average is over several realisations of the same protocol,
ending at time $t$ with energy $E(t)=E_o$.

From the Langevin point of view, the left hand side of 
(\ref{edwards4})
 corresponds to adding
over all periodic trajectories  starting from an energy $E_o$ with
period $t^*$. Equation (\ref{edwards4}) 
tells us that thermal histories give the same result as a very particular set of 
trajectories, in a manner analogous as when one represents chaotic
dynamical
systems by using only the periodic orbits.

In the zero-temperature or in the mean-field case,
 we can set $t^*=\infty$ in (\ref{edwards4}), and thus select the 
states with infinite lifetime. 
This is Edwards' prescription for granular media ($T \sim 0$) and the
one we discussed above within mean-field.
In finite dimensions, where metastable 
states eventually nucleate, we are forced to give $t^*$ a finite
value. The fact that the choice of $t^*$ is not unique already tells
us that equation  (\ref{edwards4}) will be an approximation.

{\em The central remaining question now is what is a reasonable value 
for $t^*$.}  
Indeed, giving a value of $t^*$ determines a configurational entropy
 ${\cal S}_{t^*}(E_o)$ and
an intra-state free-energy: $f_{t^*}(E_o)=E_o-T s_{t^*}(E_o)$. This in turn
determines a temperature associated with the timescale as:
\begin{equation}
T_{t^*} \equiv \left[ \frac{\partial {\cal S}_{t^*}}{\partial f_{t^*}}\right]^{-1}
 \label{teff}
\end{equation}
(where we have eliminated the energy in favour of $f_{t^*}$).
It seems now natural  to compare the different relaxation times 
of the correlations in a given problem with $t^*$. For example, glassy systems
can often be described with two timescales, a fast (`$\beta$') relaxation and a slow,
(waiting time dependent if the system is aging)  `$\alpha$-relaxation' $t_\alpha$.
If $t^*$ is small ($t^{*}<<t_{\beta }$), Eqn. (\ref{teff}) gives $T_{eff}
 \sim T$ \cite{foot}.
If we put instead:
\begin{equation}
t_{\beta }<<t^*\lesssim t_{\alpha }
\end{equation}
then $T_{t^*} $ may be different (larger) than $T$. This is the temperature to compare 
with the one governing the relation between correlation and response in the regime
corresponding to the $\alpha$-relaxation \cite{foot3}.

There are cases in which the `slow' relaxations happen in several timescales,
becoming more and more different as aging proceeds \cite{review}.
Then, the definition (\ref{teff}) immediately yields a different temperature for every 
widely separated timescale, and this would  agree with  what  
one observes from fluctuation-dissipation relations \cite{review}.

\vspace{.5cm}

{\it Time dependent order parameters}

\vspace{.5cm}

We are interested in calculating two-time correlation functions:
\begin{equation}
<A(t)B(t')> \equiv \frac{1}{\cal{N}} {\mbox{tr}} \; 
\left[ e^{-(t^*-t)H} A e^{-(t-t')H} B
  e^{-t'H} \right] = 
\frac{1}{\cal{N}}{\mbox{tr}} \;   \left[ e^{-(t^*-\tau)H} A e^{-\tau H}
  B\right] = <AB>(\tau)
\label{trace}
\end{equation}
where $\tau=t-t'$, and ${\cal{N}}$ is the normalisation.
Cyclic permutation implies:
\begin{equation}
<AB>(\tau) = <BA>(t^*-\tau)
\end{equation}

If $H$ is a time-dependent Fokker-Planck operator, associated with
forces deriving from a potential $E$, we have:
\begin{equation}
e^{\beta E} H e^{-\beta E} = H^\dag
\label{reversal}
\end{equation}
Using (\ref{reversal}) in (\ref{trace}) we get the {\em time-reversal property}:
\begin{equation}
<AB>(\tau) =  tr^{*}   \left[ e^{-(t^*-\tau)H} 
(e^{\beta E} B e^{-\beta E})^\dag
 e^{-\tau H}
 (e^{\beta E} A e^{-\beta E})^\dag ]  \right]= <{\tilde B}{\tilde
 A}>^*(\tau)
\label{ffddtt}
\end{equation}
where ${\tilde A}\equiv (e^{\beta E} A e^{-\beta E})^\dag $ and
${\tilde B}\equiv (e^{\beta E} B e^{-\beta E})^\dag $

We shall need to consider the cases in which $A$ and $B$ are respectively
 $S_i$ and 
$ \hat{S}_i \equiv -\frac{\partial}{\partial S_i}$.
Let us define, for a system of $N$ degrees of freedom:
\begin{eqnarray}
C(t-t') &=& C(\tau) = \frac{1}{N}\sum_i <S_i(t)S_i(t')>  \nonumber \\
R(t-t') &=& R(\tau) =\frac{1}{N}\sum_i < S_i(t)\hat S_i(t')>  \nonumber \\
D(t-t') &=& D(\tau) =\frac{1}{N}\sum_i < \hat{S}_i(t)\hat{S}_i(t')>
\end{eqnarray}

Because we are considering periodic trajectories, causality is
violated. This means that neither $D(\tau)$ nor $R(\tau)$ for negative
$\tau$ need to be zero. 
However, it is easy to show that if $(t^*-\tau)$ is larger than the
thermalisation time (in which the system is projected to its Gibbs
measure), then we recover $D(\tau)=0$ and $R(\tau)R(-\tau)=0$.
The existence of solutions violating causality for $t^*$ large 
is then a symptom of 
large equilibration times, i.e. of glassiness.

Using the time-reversal equation (\ref{ffddtt}), 
it is easy to derive a  non-causal form
of FDT,
valid for any $t^*$:
\begin{equation}
\frac{\partial}{\partial t'} C(t-t')=
T \left\{ R(\tau)-R(-\tau) \right\}
\label{ffddtt1}
\end{equation}

The physical meaning of $D$, unlike that of $C$ and $R$, is unfamiliar.
 If we couple
the states to a time-dependent, random magnetic  `pinning fields' \cite{Remi}
$h_i$ such that ${\overline{h_i(t) h_j (t')}}=F(t,t') \; \delta_{ij}$,
the  fields will make  a change in the number of metastable states,
and  it is easy to
 show that:
\begin{equation}
D(t,t')=\frac{1}{N}\frac{\delta}{\delta F(t,t')} 
\ln \left\{ {\mbox{tr}} [e^{-t^* H}] \right\}
\label{ddeeqq}
\end{equation}
It is hence clear that systems with a finite number of states
will have $D=0$ in the long-time limit.

\section{The calculation}

In the following we apply the theory discussed above
 to a simple glassy system: the spherical version of model
 (\ref{pspin}).
These models are thought to be mean-field versions of structural
 glasses,
we shall not deal in this paper with models corresponding to
 spin-glasses for which, as mentioned above, we do not expect the
 present computation of states to be in correspondence with the
 TAP-equation
based calculations in the literature.

The trace of the Fokker-Planck operator, at a fixed energy density $E$, can 
be written as a functional integral \cite{ZJ}
over the spin fields $S_i (t)$ and the response fields $\hat{S}_i(t)$
with periodic boundary conditions on $S_i (t)$.
Once the average of the trace has been performed \cite{foot1},
 the action depends
on the fields $S_i (t)$ and $\hat{S}_i(t)$ through the 
two time functions $C(t,t'),R(t,t'),D(t,t')$ only. As a consequence
one can integrate out the fields $S_i (t)$ and $\hat{S}_i(t)$ and
get an effective action on two time functions:
\begin{eqnarray}\label{action}
S/N&=&-\int_{0}^{t^{*}}dt\left.\left(\partial_{t}R(t,t')
+\lambda R(t,t')-TD(t,t')\right)\right|_{t'=t^+}\\
&+&\frac{p}{4}
\int_{0}^{t^{*}}dtdt'\left(D(t,t')C^{p-1}(t,t')+(p-1)R(t,t')R(t',t)
C^{p-2}(t,t')\right)\nonumber\\
&-&\frac{\hat \lambda }{2}\int_{0}^{t^{*}}dt\left( C(t,t)-1 \right)+
\frac 1 2\mbox{Tr}\ln \mbox{M}\label{e1.2}\nonumber
\end{eqnarray}
where the operator $M$ reads:
\begin{eqnarray}
\label{M}
M= 
\left( \begin{array}{cc} 
R(t,t')& C(t,t')\\
D(t,t')& R(t',t)
\end{array} \right) \qquad ,
\end{eqnarray}
Since we consider times of order one with respect to N the functional
 integral is dominated by a saddle point contribution. 
We shall obtain periodic dynamic solutions which, in the glassy phase
  {\em (a)} break causality, {\em (b)} have non-zero action,
{\em (c)} satisfy 
time-translational invariance, and {\em (d)} satisfy time-reversal
and its consequence (\ref{ffddtt1}).
Note that {\em (a)} and {\em (b)} are properties typical
of instantons, while {\em (c)} and {\em (d)} are not.
In the high temperature phase there is a periodic solution with
 zero action for long times corresponding essentially to 
 the equilibrium dynamics.

The
 stationarity conditions on the action are equivalent to four
 equations on the two-time functions:
\begin{equation}
C'(\tau)=-\lambda C(\tau)+ 2T R(-\tau)  
+\frac{p}{2}\int_{0}^{t^{*}}dt'' C^{p-1}(t-t'') R(t'-t'') 
+k \int_{0}^{t^{*}}
R(t-t'') C^{p-2}(t-t'') C(t''-t') dt''
\label{c-eq}
\end{equation}

\begin{equation}\label{r-eq1}
R'(\tau )=-\lambda R(\tau )+ 
2T D(\tau) + \frac{p}{2}\int_{0}^{{t^{*}}}dt''C^{p-1} (t-t'') D(t'-t'')+ 
 k \int_{0}^{t^{*}}dt'' C^{p-2} (t-t'') R(t-t'')R(t''-t')+\delta (\tau )
\end{equation}
\begin{eqnarray}\label{r-eq2}
R'(\tau )&=&-\lambda R(\tau )+ 
k\int_{0}^{{t^{*}}}dt''D(t'-t'')C^{p-2}(t'-t'') C(t-t'')
+ k  \int_{0}^{t^{*}}dt''C^{p-2}(t'-t'')R(t-t'')R(t''-t')\nonumber\\
&+&k (p-2)\int_{0}^{t^{*}}dt''C^{p-3}(t'-t'')R(t'-t'')R(t''-t')C(t-t'')
-\hat {\lambda }C(t-t')+\delta (\tau )
\end{eqnarray}

\begin{eqnarray}\label{d-eq}
-D'(\tau )&=&-\lambda D(\tau )+
k \int_{0}^{t^{*}}dt''D(t'-t'')R(t''-t)C^{p-2}(t-t'')
+k\int_0^{t^*}dt''D(t-t'')C^{p-2}(t-t'')R(t''-t')
\nonumber \\
&+&k (p-2)\int_{0}^{t^{*}}dt''R(t-t'')R(t''-t)R(t''-t')C^{p-3}(t-t'')
-\hat {\lambda }R(\tau )\nonumber
\end{eqnarray}
where $k=\frac{p(p-1)}{2}$ and we make explicitly use of the time translation
 invariance  $\tau=t-t'$.
The spherical condition fix the value of $\hat \lambda $, which can be 
obtained subtracting eq. (\ref{r-eq2}) to eq. (\ref{r-eq1}) for $\tau =0$:
\begin{equation}\label{lambdac}
\hat{\lambda}=(p-2)\left( p/2\int_{0}^{{\cal
\tau}}dt''
C^{p-1}(t-t'')D(t-t'')
+ k \int_{0}^{{\cal \tau}}dt''R(t-t'')
R(t''-t)C^{p-2}(t-t'')\right)-2TD(0)
\end{equation}
Moreover fixing the value of the energy $E$ gives an equation on
the spherical multiplier $\lambda$:
\begin{equation}\label{lambda}
pE=-\lambda+T(R(0^+)+R(0^-))
\end{equation}
as a consequence $E$ and $\lambda$ are directly related. 
Using the FDT relation one can show that 
(\ref{c-eq}-\ref{d-eq}) reduce to a set of three independent equations on the
functions: $C(\tau )$, $R(\tau)+R(-\tau)$ and 
$D(\tau )$. 
\subsection{Time-reversal, non-causal solutions}\label{zf}
Let us  show that computing for very large $t^{*}$ 
 the trace of the Fokker-Planck operator one can recover the 
 the number of stable states, and the dynamics within these states.
 The number of stable states can be obtained by a pure static computation 
 for the p-spin spherical model ($p>2$) using the TAP
 equations \cite{crisomtap}.

For very large 
$t^{*}$ there are two possible behaviours for the two-time functions 
 depending on the energy (and the model) we consider: 
 
\begin{itemize}

\item if at the  energy value considered there are  stable states then
the action evaluated in the solution has a well defined limit as
$t^* \rightarrow \infty$. In this same limit, 
 one expects that the two-time
functions  for finite $\tau$ 
 describe the dynamics inside a stable state 
(as calculated previously with other methods \cite{Bamebu}).
 A careful analysis of equations
 (\ref{c-eq}-\ref{d-eq}) allows one to show that the asymptotic forms
 of two-time quantities reads for $\tau <<t^{*}$ :

\begin{eqnarray}\label{asymp}
C(\tau)&=&C_c(\tau )+\frac{1}{t^{*}}\hat C(\tau ), \qquad R(\tau)=
R_c(\tau)+\frac{1}{t^{*}}(\hat R(\tau )+ r-r_c) ,\\
 \label{asymp1}
D(\tau)&=&\frac{1}{t^{*}}D_o(\tau)
+\frac{1}{(t^{*})^{2}}\hat D\left(\frac{\tau}{t^*} \right)
\end{eqnarray}
The function $R_c(\tau)$ is causal, and for large $\tau$ (but small
with respect to $t^*$) we have that
$R_c(\tau) \rightarrow 0$ and  $C_c(\tau) \rightarrow q$. 
Together $R_c$ and $C_c$  describe the relaxation within a state.
All other functions are of order one when their arguments are of order
one, and tend to zero when their argument is large.
 The corrections of order  $1/t^{*}$ are {\em not} subleading 
 in the  computation of the trace, as this involves integrals
over a large  interval $[0,t^*]$.
The Edwards-Anderson parameter  $q$ and $r-r_{c}$ are order one constants 
to be determined in what follows.
 Note that the scaling of $D$ implies that $\hat \lambda $
 is of  order  $1/t^{*}$.
 One can easily check that $C_c(\tau)$ and $R_c (\tau)$ satisfy
 the equations describing the equilibrium dynamics inside a state
 studied in \cite{Bamebu}:
\begin{eqnarray}
C_c'(\tau) &=& \lambda  C_c(\tau) + p/2 
\int_{0}^{{t^*}}dt'' C^{p-1}_c(t-t'') R_c(t'-t'') \nonumber \\
&+&k \int_{0}^{{\rm L}}
R_c(t-t'') C_c^{p-2}(t-t'') C_c(t''-t') dt''+ p^2q^{p-1} (r-r_c)/2
\label{c-eq-fast}
\end{eqnarray}
 $R_c(\tau )$ is causal
 and related to the correlation function through the FDT relation
$R_c(\tau)=-1/TC_{c}'(\tau)\theta(\tau )$.

\item if for the energy value considered there are no stable states then
 the behaviour of the two-time functions is similar to the  previous one 
 except that $D$ has  a part of  order of one for finite time $D_o$ 
 and a part of  order or $1/t^{*}$ for $\tau \sim t^{*}$:
\begin{equation}
D(\tau)=D_o(\tau)
+\frac{1}{(t^{*})}\hat D\left(\frac{\tau}{t^*} \right)
\end{equation}
As a consequence $C(\tau)$ and $R(\tau)$  do not satisfy 
equations `within a state', and one has to solve a set of 
 three equations on $C$, $R$ and $D$ in which $R$ is not causal
 also for infinite times ($t^{*}$).   
\end{itemize}

In the rest of this subsection  we consider an energy such that stable states
exists and we compute the zero-frequency values of $C$, $R$ and $D$.
Using the asymptotic form introduced before one finds:
\begin{eqnarray}\label{int1}
\int_{0}^{t^{*}}C(\tau)d\tau &=&qt^*+O(1)\\
\int_{0}^{t^{*}}R_c(\tau)d\tau &=&r_{c}=\frac{1-q}{T}\label{int3}\\
\int_{0}^{t^{*}}R(\tau)d\tau &=&\int_{0}^{t^{*}}R_c(\tau )
+r-r_{c}+O(1/t^{*})=r+O(1/t^{*})\label{int2}\\
\int_{0}^{t^*}D(\tau)d\tau &=&\frac{d}{t^{*}}+O(1/(t^{*})^{2})\label{int4}
\end{eqnarray} 
Moreover for large $t^{*}$ the relation (\ref{lambda}) reduces to
the usual one: $pE=-\lambda +T/2+O(1/t^{*})$. Therefore 
we can consider $\lambda$ as a fixed parameter.\\
Integrating eq. (\ref{c-eq}) between $0$ and $t^*$ and taking the
leading order in $t^*$ we obtain:
\begin{equation}\label{la0}
q\left(-\lambda+ \frac{p}{2T}(1-q)q^{p-2}  + 
  \frac{p}{2T}(1-q^{p-1})+
\frac{p^2}{2}(r-r_c) q^{p-2} \right)=0
\end{equation}
Subtracting eq. (\ref{c-eq}) evaluated in $\tau =0$ to (\ref{la0}) we
get the usual equation on $\lambda$
and $q$:
\begin{equation}\label{la1}
\lambda=\frac{T}{1-q}+\frac{p}{2T}(1-q^{p-1})
\end{equation}
Since the spherical multiplier is a parameter, this equation fixes the
overlap. It is useful to write this equation in a way which is
directly related to the static computation. Using the following
notation:
\begin{equation}\label{la2}
  q^{p/2-1}p{\cal E}=-\lambda
  +\frac{p}{2T}(1-q^{p-1})-\frac{p(p-1)}{2T}q^{p-2}(1-q)
  \qquad \qquad z=(1-q)q^{p/2-1}/T
\end{equation}
where ${\cal E}$ corresponds to the zero-temperature or radial energy
which appear in the static computation, one can rewrite (\ref{la1})
as the usual static equation on $q$ \cite{crisomtap}
\begin{equation}\label{z}
1+pz{\cal  E}+p(p-1)z^2 /2=0
\end{equation}
From equation (\ref{la0}) we find the value of $r$:
\begin{equation}\label{r}
r=-\frac 2 p q^{1-p/2}{\cal E}
\end{equation}
Integrating $(\ref{r-eq1})$ between $0$ and $t^{*}$ and using eq.
(\ref{la0}) we get the value of $d$:
\begin{equation}\label{d}
q^{p-1}d=-2/p+4{\cal E}^{2}/p^{2}
\end{equation}
Finally we confirm,  using the relationship
between $\lambda$ and $q$,  that 
the equations on $C_c(\tau)$ and $R_c(\tau)$ (\ref{c-eq-fast}) 
 are indeed the same ones found in \cite{Bamebu} for the relaxational
dynamics inside the stable state with overlap $q$ and energy 
$E=(-\lambda +T/2)/p$.
\subsection{The configurational entropy}

To obtain the number of stable states we have to inject the 
 solution of eqs. (\ref{c-eq}-\ref{d-eq}) into the action (\ref{action})
 and then take the long time limit. Using the compact notation $Q(\tau ;
t^{*})=(C(\tau ; t^{*}),R(\tau ; t^{*}),D(\tau ; t^{*}))$
 for the set of the two-time functions, we decompose the asymptotic solution 
as $Q(\tau ;t^{*})=Q_{0}(\tau ;t^{*})+Q_{1}(\tau ;t^{*})/t^{*}$,
where $Q_{0}$ reads:
\begin{equation}\label{Q0}
Q_{0}(\tau ;t^{*})=\left(C_c(\tau),R_c(\tau)+\frac{r-r_{c}}{t^{*}},
\frac{d}{(t^{*})^{2}} \right)
\end{equation}
and $C_c(\tau ),R_c(\tau )$ are the solution of the relaxational dynamics
 inside a stable state (\ref{c-eq-fast}) with the values of $r,r_{c},d$ 
 determined above.
 Using this decomposition and the fact that $Q$ is a saddle point of 
(\ref{action}) we find for very large $t^{*}$:
\begin{equation}\label{actionasymp}
S(Q)\simeq S(Q_{0})-\frac{1}{2}\frac{Q_{1}}{t^{*}}
\otimes \frac{\delta ^{2}S}{\delta Q^{2}}\otimes \frac{Q_{1}}{t^{*}} + ...
\qquad t^{*}>>1
\end{equation}
An explicit computation shows that the second term of (\ref{actionasymp})
vanishes in the long time limit provided that the corrections
 to $q$ are of an order less than $1/t^{*}$. This seems natural to us
 since we are considering the relaxation dynamics inside a stable state.   
As a consequence the (annealed average of the) 
logarithm of the number of stable states
 coincides with $S(Q_{0})$. When we inject $Q_{0}$ into (\ref{action})
 the first two lines can be easily computed and read:
\begin{equation}\label{twolines}
-\lambda (r-r_{c})+\frac{p}{4}dq^{p-1}+\frac{p}{2T}(r-r_{c})(1-q^{p-1})+
\frac{p(p-1)}{4}(r-r_{c})^{2}q^{p-2}
\end{equation}
whereas the computation of the third one, which reduces only to 
 Trln$M/2$, is slightly more subtle. Since the operator (\ref{M})
 is diagonal in Fourier space, we get:
\begin{eqnarray}
\label{trace1}
\frac{1}{2}\mbox{Tr}\ln M= \frac{1}{2}\ln (r^{2}-qd)+
\frac{1}{2}\sum_{\omega \neq 0}\ln
\left( \begin{array}{cc} 
\hat R (\omega )& \hat C(\omega )\\
0& \hat R^{*}(\omega )
\end{array} \right) \qquad ,
\end{eqnarray} 
where the Fourier transform of a function $F(\tau )$ is defined as:
\begin{equation}\label{Fourier}
\hat F(\omega )=\int_{0}^{t^{*}}e^{-i\omega \tau }F(\tau )d\tau 
\qquad ,\qquad \omega =\frac{2\pi n}{t^{*}}
\quad n=0,\pm 1,\pm 2, \dots 
\end{equation}
The  function $R_c(t,t')$ is causal and the associated operator
 is upper triangular with
diagonal elements equal to unity.  Its  determinant hence is one
(here the Ito convention
is crucial), as it should because it does not 
give any contribution to the action in the standard case. As a consequence
we have:
\begin{eqnarray}
\label{det2}
\frac{1}{2}\sum_{\omega\neq 0}\ln
\left( \begin{array}{cc} 
\hat R (\omega )& \hat C(\omega )\\
0& \hat R^{*}(\omega )
\end{array} \right)=-\ln (r_{c}) \qquad ,
\end{eqnarray} 
Collecting all the pieces together and using the equation on $q,r,r_{c}$ and
$d$ obtained in section \ref{zf} we find that the action $S(Q_{0})$ reads:
\begin{eqnarray}\label{sp}
  S(Q_{0})/N&=&\frac{1}{2}(1+\ln p-\ln 2)-{\cal E}^2+\frac{1}{2}
\left(\frac{{\cal E}- \sqrt{{\cal E}^2-{\cal E}_c
        ^2}}{{\cal E}_c}\right)^2
  +\ln (-{\cal E}- \sqrt{{\cal E}^2-{\cal E}_c ^2})\\
{\cal  E}_{c}&=&-\sqrt{\frac{2(p-1)}{p}}\nonumber
\end{eqnarray}
As expected, this expression coincides with the logarithm of the number  
of TAP states computed by Crisanti and Sommers \cite{crisomtap}.

Note that this formula is correct only for ${\cal E}<{\cal E}_{c}$. For
${\cal E}>{\cal E}_{c}$ the formalism tells us that there
are  no stable states as follows: in this energy regime
$D$ and $\hat \lambda $ remain of  order  one even for  infinite
 $t^{*}$ and the action acquires a  negative contribution  of  order 
 $t^{*}$. This is as it should: since for these energy values there are
 only metastable states with finite life-times, the longer we set
 $t^*$
the less metastable states we find.

In general for finite dimensional glassy systems 
 the interesting quantity will be the logarithm of the 
 number of metastable states with finite lifetime,
 which can be obtained plugging  the 
 solution $Q(\tau)$ into the action S for a finite value of 
 $t^{*}$. In the following we compute this quantity at zero
 temperature.

\subsection{The zero temperature case}\label{T0.sec}
\begin{figure}[bt]
\centerline{    \epsfysize=8cm
       \epsffile{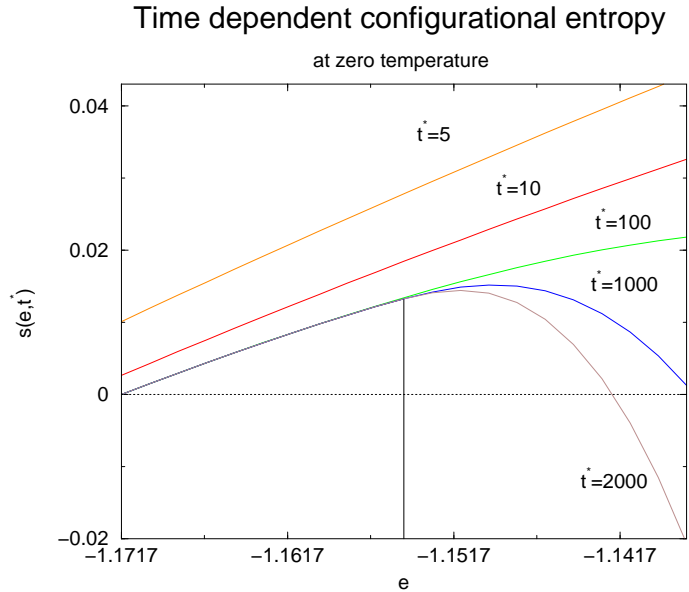}}
\caption{Time dependent configurational entropy $s(e,t^{*})$ 
at $T\rightarrow 0$
for $p=3$ as a function of the energy density $e$ and of 
the time $t^{*}$. From top to bottom: 
$t^{*}=5,10,100,1000,2000,+\infty$. Note the infinite time limit
curve with positive configurational entropy {\it a mean-field artifact}. 
\label{T0.fig}}
\end{figure}

At zero temperature the equation (\ref{r-eq1}) is particularly simple.
Introducing  the notation $R(\tau )=R_{c}(\tau )+(r-r_{c})/t^{*}$
we find that the equation on the Fourier transform of $R_{c}$ reads:
\begin{equation}\label{rT0}
-i\omega \hat R_{c}(\omega)-\lambda \hat R_{c}(\omega)+\frac{p(p-1)}{2}
\hat R_{c}(\omega )\hat R_{c}(\omega )+1=0
\end{equation}
For each frequency there are two solutions:
\begin{equation}\label{2sol}
\hat R_{c}^{\pm}(\omega)=\frac{1}{p(p-1)}\left(-\Lambda \pm \sqrt{\Lambda ^{2}
-2p(p-1)}\right),\qquad \Lambda =\lambda +i\omega
\end{equation}
 In the following we focus on the two solutions $R_{c}^{+}(\tau )$ 
and $R_{c}^{-}(\tau )$ which correspond respectively 
to taking the Fourier transform of $\hat R_{c}^{+}(\omega )$ and
$\hat R_{c}^{-}(\omega )$. Using that at zero temperature $C(\tau )=1$
 one can decompose $S(Q)$ in two terms such that all the dependence
on $t^{*}$ is contained only in one of them:
\begin{eqnarray}\label{T0sp}
  S/N(Q^{\pm})&=&\frac{1}{2}(1-\ln p-\ln 2)-{\cal E}^2+\frac{1}{2}
\left(\frac{{\cal E}\mp \sqrt{{\cal E}^2-{\cal E}_c
        ^2}}{{\cal E}_c}\right)^2
  +\ln (-{\cal E}\mp \sqrt{{\cal E}^2-{\cal E}_c ^2})\\
  &-&\frac{p(p-1)}{4}
\int_{0}^{t^{*}} R_c^{\pm}(\tau )R_c^{\pm}(-\tau )d\tau 
  +\frac{1}{2}\sum_{\omega } \ln (\hat R_c ^{\pm}(\omega )
  \hat R_c ^{\pm}(-\omega ))\nonumber
\end{eqnarray}
The computation of the second line is performed in the Appendix B.
It turns out that, as in the static case \cite{Cagipa}, the dominant 
contribution is given by $R_{c}^{-}$ for 
${\cal E}_{RSB}<{\cal  E}<{\cal E}_{c}$, 
by $R_{c}^{+}$ for $-{\cal E}_{c}<{\cal  E}<-{\cal E}_{RSB}$ 
 and for $-{\cal E}_{c}<{\cal  E}
<-{\cal E}_{c}$ the two saddle point contributions are the same
 (see Appendix B).
Note that at zero temperature, 
${\cal E}$ is the energy density of the system.
The final result is 
\begin{eqnarray}\label{Sfin}
S(Q)&=&\mbox{Re}\left(\frac{1}{2}(1-\ln p-\ln 2)-{\cal E}^2+\frac{1}{2}
\left(\frac{{\cal E}\mp \sqrt{{\cal E}^2-{\cal E}_c
        ^2}}{{\cal E}_c}\right)^2
  +\ln (-{\cal E}\mp \sqrt{{\cal E}^2-{\cal E}_c ^2}) \right)\\
&-&\int d\omega \rho_p(\omega+p{\cal E})
\ln (1-\exp(-t^{*}|\omega|))+t^{*}\int_{-\infty}^{0} d\omega \rho_p(\omega+p
{\cal E})\nonumber
\omega 
\end{eqnarray}
where $\rho_p(x)=\sqrt{2p(p-1)-x^2}/(\pi p (p-1))$ is the Wigner semi-circle
law. In fig. \ref{T0.fig} we plot (\ref{Sfin}) for $p=3$ and different values
 of $t^{*}$ as a function of the energy density ${\cal E}$. For very large 
values of $t^{*}$, $S(Q)$ converges to the logarithm of the number of stable 
states. Note that for a finite dimensional system we expect a similar behavior
but with a vanishing curve for infinite $t^{*}$. Finally, we remark
that the formula (\ref{Sfin}) has a simple interpretation. In fact the
first line coicides with the number of saddles with energy density
${\cal}E$. Moreover, since
the spectrum of the Fokker-Planck operator for an harmonic oscillator 
with frequency $\omega $ is \cite{ZJ} $E_{n}=(n+1/2)|\omega |-\omega
/2$, the second line of (\ref{Sfin}) corresponds to the contribution
due to a collection of harmonic oscillators with frequency distributed
by the semicircle law centered in $-p{\cal E}$. This distribution
is exactly the same of the eigenvalues of the energy Hessian
 evaluated in saddles with energy density ${\cal E}$ \cite{Cagipa}.
As a consequence, at zero temperature, the spectrum of the
Fokker-Planck operator for the p-spin spherical model coincides with
the one obtained making an harmonic expansion around each saddle
(also the instable ones).
\section{The two-groups Ansatz and supersymmetry breaking}
Twenty years ago, when people started to search for replica symmetry
breaking solutions of the Sherrigton-Kirkpatrick model,
Bray and Moore \cite{twogroups1} proposed a two-groups Ansatz
for the famous $Q_{a,b}$ matrix \cite{Beyond}. At a first sight the
results seemed a little bit strange since
the $\lim_{n\rightarrow 0}\overline{Z^n}\neq 1$ ! ($\overline{\cdot}$
means the average over disorder). It turned out \cite{twogroups2} 
that in the limit
of $n$ going to zero, the logarithm of $\overline{Z^n}$ equals the
the logarithm of the number of TAP states, i.e. the long-time limit
 of the configurational entropy (\ref{compl}).
The reason of this ``coincidence'' has been completely obscure until
now. For instance, for many mean-field spin glass models the
complexity was computed starting from TAP states and, after, it was
checked that the two-groups Ansatz gave back the correct result
\cite{twogroups2,potters,enfants}.\\
Using the properties of the dynamical solutions presented in previous
sections we can unveil why the Bray and Moore Ansatz allows one to
calculate the long-time limit of the configurational entropy (\ref{compl}). 
In fact this Ansatz is {\em isomorphic} to zero frequency
part of 
the dynamical calculation for $t^*\rightarrow \infty$. This can be
shown by the supersymmetric formalism for Langevin dynamics
\cite{ZJ,Kurchansusy}. Within this framework all the two points
correlation functions between fields $S_i(t)$ and $\hat{S}_i(t)$
can be encoded in:
\begin{eqnarray}\label{susy1}
  Q(1,2)&=&\sum_{i=1}^N \left<S_i (1) S_i (2) \right>\qquad 1=(Tt_1,\overline{\theta}_1,\theta _1)\\
  S_i(1)&=&S_i(t_1)+\overline{\theta}_1\theta _1\hat{S}_i(t)+
  \overline{c}_i(t_1)\theta_1+\overline{\theta}_1 c_i(t_1) \label{susy2}
\end{eqnarray}
where $\overline{\theta}_1,\theta _1$ are Grassmann variables and
$\overline{c}_i(t),c_i (t)$ are fermion
fields \cite{ZJ,Kurchansusy}.  
Using this formalism the dynamical solution (\ref{Q0}) giving back the
configurational entropy reads at large times:
\begin{eqnarray}\label{susy3}
  Q(1,2)&=&Q_c (1,2)+q+(\overline{\theta}_1\theta _1+\overline{\theta}_2
  \theta _2)\frac{r-r_c}{t^*}+\overline{\theta}_1\theta _1\overline{\theta}_2
  \theta _2\frac{d}{(t^*)^2}\\
  Q_c (1,2)&=&\left(1+\frac{1}{2}(\overline{\theta}_{1}-\overline{\theta}_{2})
[\theta_{1}+\theta _{2}-(\theta_{1}-\theta _{2}){\mbox{sign}}(t_{1}-t_{2})]
\frac{1}{T}\frac{\partial}{\partial t_{1}} \right)
(C_c(t_1-t_2)-q)\label{susy4}
\end{eqnarray}

The function $Q_c(1,2)$ is supersymmetric, whereas the last
two terms in the right hand side of (\ref{susy3}) break supersymmetry.\\
On the other hand, the two-groups Ansatz consists in a
symmetric $Q_{a,b}$ matrix:
\begin{eqnarray}\label{twogroups1}
  Q_{a,b}&=&1+\frac{B}{m} \qquad a=b\leq m,
  \qquad Q_{a,b}=1-\frac{B}{m} \qquad m<a=b\leq n\\\label{twogroups2}
  Q_{a,b}&=&A+\frac{B}{m} \qquad a\neq b;\quad a,b\leq
  m\\\label{twogroups3}
  Q_{a,b}&=&A-\frac{B}{m} \qquad a\neq b;\quad m<a,b\leq
  n\\\label{twogroups4}
  Q_{a,b}&=&A-\frac{C}{m^2} \qquad a\leq m, \quad m<b\leq n
\end{eqnarray}
where one has to take the
$m\rightarrow \infty$ and $n\rightarrow 0$ limits.
The functional dependence of the dynamical and replica free
energy of, respectively, $Q(1,2)$ and $Q_{a,b}$ is the same
\cite{Kurchansusy}. Indeed, the kinetic term in the dynamical free
energy, which does not have a correspondent in the static case,
is zero for the dynamical
solution (\ref{susy3}). Moreover if one puts $B=(r-r_c)T$, $2C=T^2d$,
$A=q$ the two matrices $Q(1,2)$ and
$Q_{a,b}$ lead to the same results under tracing, convolution and
term by term product. For instance, one can easily obtain:
\begin{eqnarray}
  \sum_a f(Q_{a,a})&=&\int d1 f(Q(1,1))=2B\\
  \sum_{a,b}f(Q_{a,b})&=&\int d1 d2
  f(Q(1,2))=2(f'(1)-f'(A))B+f''(A)B^2+2f'(A)C\\
  \mbox{{Tr}}_{a,b}\log Q&=&{\mbox{Tr}}_{1,2}\log Q=-2\log (1-A)+\log((1-A+B)^2-2AC)
\end{eqnarray}  
As a consequence the computation, which make use of the two-groups
Ansatz, is {\em isomorphic} to the dynamical
one for $t\rightarrow t^*$. Therefore the replica symmetry breaking
scheme encoded in this Ansatz can be finally understood: it is a
way to implement the dynamical computation in a replica formalism.
There are, however, two important differences between the two approaches.
First of all, in the dynamical computation we are not free to choose
between different Ans{\"a}tze the one which gives back the 
 long-time limit of the configurational entropy (\ref{compl}),
but we have simply to solve the equations of motion. This clearly makes the
procedure inambiguous,  unlike the case of  the replica computation.
Moreover the two approaches lead to the same results only if
a dynamical solution with the correct values of $q,d,r$ exists.
It could then happen (cf. the discussion on the configurational entropy of the SK
model) that the equations on $q,d,r$ admit a solution but
there is no dynamical solution corresponding to these values.
As a consequence even if the static computation, i.e. the
sum over all TAP solutions or the computation by the two-groups
scheme, predicts the existence of an exponential number of stable states
 the more correct dynamical calculation does not.

\section{Conclusions}
In this work we have shown how to put the questions related 
 to metastable states in glasses in a  manner valid for 
finite-dimensional systems.
 
 We have used the construction of Gaveau and Schulman to define the 
metastable states. This construction requires the
existence of a `gap' in the lifetime, so that one can associate  `states' 
with  distributions that are stable for much longer than a given time $t^*$,
and `transient processes' those that decay  much  faster.
There is no such gap in real glasses, so our use of this construction
has to be considered partly as a definition inspired in the cases where
 there is.  

A reasonable criterion for the relevance of any quantity will hence be  
 that they are not too sensitive to
the  exact value chosen for $t^*$. For example, if we consider a
temperature associated with the $\alpha$-relaxation as $T_{t^*}$ with
$t^* \stackrel{<}{\sim} t_\alpha$, this definition is meaningful to
the extent that it is stable with respect to a change in $t^*$ of, say,
an order of magnitude.

Next, there is the question as to whether $T_{t^*}$ indeed reproduces
the fluctuation-dissipation temperature. This and other results depend
on the validity of the flatness hypotheses {\`a} la Edwards (for which positive
 evidence begins to appear \cite{aggiunta}).
In this paper we have formulated the hypothesis in a manner applicable
to positive temperatures and finite dimensions (as well as to vibrated
systems)---  but we have {\em not} attempted to prove it.
It may be, however, that writing it in the form (\ref{edwards4}), can be a good
starting point for doing this.
Moreover, the form  ({\ref{edwards4}}) (and (\ref{teff})) 
lends itself naturally to a
 generalisation to  cases in which a
system has more than two widely separated timescales and temperatures.

Finally, the computation in section V has allowed us to check the
mean-field results without relying on the  TAP states, themselves an
intrinsically mean-field concept.
The kind of solutions that dominate break causality and have positive
action, but satisfy time-reversal and a non-causal form of FDT.
Unlike the barrier-crossing solution of Lopatin and Ioffe \cite{Ioffe}, 
they have in this sense only some of  the properties of true instantons.
Moreover, the dynamical computation unveils
the meaning of the two-groups Ansatz \cite{twogroups2}, which allows one 
to compute the number of stable states within a replica 
formalism. 

\vspace{3cm}

{\bf Acknowledgments}

 We wish to thank L. Ioffe, 
J. L. Lebowitz and L. S. Schulman for useful suggestions.

\pagebreak

\section{Appendix A. Pure Barriers.}

Let us show how to define `pure barriers' using a supersymmetric
extention 
of the Fokker-Planck operator.

For a system with $N$ degrees of freedom, introduce the $N$ fermion
creation and annihilation operators $a_i$ and $a^\dag_i$, with
anticommutation relations $[a_i,a^\dag_j]_+=\delta_{ij}$.
Define the supersymmetric charges as:
\begin{equation}
 {\bar Q}= (Tp_i-iE_{,i})a^\dag_i \;\;\; ; \;\;\; Q=p_ia_i
\end{equation}
The supersymmetric operator $H_{SUSY}$
\begin{equation}
H_{SUSY}= [{\bar Q},Q]_+ = H + \frac{\partial^2 E}{\partial x_i \partial x_j}
 a^\dag_j a_i 
\end{equation}
commutes with the charges, and with the fermion number operator.
In the zero-fermion subspace it coincides with the Fokker-Planck
operator.

Applying the operator $ {\bar Q}$ to all but the lowest 
 Fokker-Planck (zero-fermion)
eigenvectors, one obtains a one-fermion eigenvector.
Going back to the example of low-temperature dynamics in a potential
of section III, one can see that the lowest one-fermion eigenvectors
 correspond to distributions associated to the barriers.
Indeed, one can convince oneself that in a low temperature
 multi-dimensional
system, the lowest eigenvectors with $k$ fermions correspond to
barriers with $k$ unstable directions (see Ref. \cite{Witten}, where
 these questions are discussed in detail, and used to derive Morse
 theory results).   

Now, by analogy with the argument motivating the construction of pure
states,
it is reasonable to define `barrier distributions' in general as 
the lowest eigenvectors of $H_{SUSY}$ with lowest eigenvalues in
subspaces
with $k$ fermions.
This definition  will make sense whenever there is a gap in the
spectrum,
whatever the origin of such a gap (large $N$, low $T$, etc).
One can then use traces of  $e^{-t^* H_{SUSY}}$ to calculate
expectation values.

\pagebreak

\section{Appendix B.}\label{T0.app}
The aim of this section is to calculate the last line of (\ref{T0sp}).
Instead of making the computation by brute force, we will use the 
exact results that can be obtained in the $p=2$ case. For $p=2$
the spherical model is a simple collection of harmonic oscillators
with frequency distributed with a semicircle law centered around
 the value of the spherical multiplier. The spectrum
 of the Fokker-Planck operator for an harmonic oscillator 
 with frequency $\omega $ is \cite{ZJ}:  
$E_{n}=(n+1/2)|\omega |-\omega /2$. Therefore for $p=2$ and at zero 
 temperature, the logarithm of the trace of the evolution operator
 reads:
\begin{equation}\label{tracep=2T=0}
\overline{\ln (Tr e^{-t^{*}H})}=-N\int d\omega \rho(\omega+2{\cal  E})
\ln (1-\exp(-t^{*}|\omega|))+N\int_{-\infty}^{0} d\omega \rho(\omega+2{\cal E})
t^{*}\omega
\end{equation}
where $\rho (x)$ is the Wigner semicircle law. One can 
obtain this result also by the functional formalism. In the computation of
section (\ref{T0.sec}) we obtained an action which for $p=2$ reads:
\begin{eqnarray}\label{Sp=2T=0}
  S/N&=&\frac{1}{2}-{\cal E}^2+\frac{1}{2}({\cal E}\mp 
  \sqrt{{\cal E}^2-1})^2+\ln(-{\cal E}\mp\sqrt{{\cal E}^2-1})\\
  &&-\frac{1}{2}\sum_n R^{\pm} (n)R^{\pm} (n)
  +\frac{1}{2}\sum_n \ln (R^{\pm}(n)R ^{\pm}(-n))\nonumber
\end{eqnarray}
where we have written $R(t)$ instead $R_{c}(t)$ since $r=r_{c}$ for $p=2$.
This is directly related to the fact that there is no configurational entropy
 for $p=2$
 and therefore the two-time 
functions relax asymptotically faster than $1/t^{*}$.
For ${\cal  E}<-1$ the $p=2$ model is a collection of stable harmonic 
oscillators. One can easily compute $R_{c}(t)$ starting from the result
 for a single oscillator and integrating over the Wigner distribution.
As expected, this function coincides with $R_{c}^{-}(t)$. As a consequence
 for ${\cal  E}<-1$ we expect that $\overline{\ln \mbox{Tr}e^{-t^{*}H}}
=NS(Q^{-})$.
On the other hand for ${\cal  E}>1$ the $p=2$ model is a collection of unstable 
harmonic oscillators which can mapped to the previous case changing the
 sign of each $\omega $. As a consequence for 
${\cal  E}>1$ we expect that $\overline{\ln \mbox{Tr}e^{-t^{*}H}}=NS(Q^{+})$.
In the intermediate energy regime a priori one has to consider both
 solutions. Since the functional computation should give back
 the result (\ref{tracep=2T=0}), the last line
 of (\ref{Sp=2T=0}) reads: 
\begin{eqnarray}\label{magic}
&&-\frac{1}{2}\sum_n R^{\pm} (n)R^{\pm} (n)
  +\frac{1}{2}\sum_n \ln (R ^{\pm}(n)R^{\pm}(-n))=\\
&=&-\int d\omega \rho(\omega+2{\cal  E})\left(-t^{*}\frac{\omega }{2}+
 \ln  |e^{\omega t^{*}/2}-e^{-\omega t^{*}/2}|\right)
 \pm i\pi \int_{-\infty}^0 d\omega \rho(\omega+2{\cal E})\nonumber
\end{eqnarray}
if the determination of the logarithm in the first line of (\ref{Sp=2T=0})
 is such that $\ln -1=-i\pi $. Note that
 for ${\cal E}<-1$ all the oscillators are stable, as a consequence
 the last term in (\ref{magic}) vanishes and (\ref{Sp=2T=0})
 coincides with (\ref{tracep=2T=0}). For ${\cal E}>1$ 
 all the oscillators are unstable, as a consequence
 the last term in (\ref{magic}) equals $i\pi $ and cancels
  the $-i\pi$ coming from the first logarithm in (\ref{Sp=2T=0}), 
 and one obtains 
 the same results that for ${\cal E}<-1$ with an additional term
 $\int d\omega \rho(\omega+2{\cal  E})t^{*}\omega $. Finally, in the 
intermediate energy regime the first line in (\ref{Sp=2T=0}) is complex
 and its imaginary part cancels exactly the imaginary contribution
 coming from the last term in (\ref{magic}). 
In this case the the two saddle point 
 contributions are the same, therefore to obtain (\ref{tracep=2T=0})
 one can consider only one of them. However to obtain the expected
 value of $R(t)$ one has to sum on saddle points.\\
For $p$ greater than two, the equation on $R_{c}$ has the same form of  
(\ref{r-eq1}) but for a $p=2$ spherical model 
 at zero temperature with a variance 
 of the couplings $J^{2}=p(p-1)/2$. As a consequence the last
 line of (\ref{T0sp}) reads:
\begin{equation}\label{lastline}
-\int d\omega \rho(\omega+p{\cal  E})\left(-t^{*}\frac{\omega }{2}+
 \ln  |e^{\omega t^{*}/2}-e^{-\omega t^{*}/2}|\right)
 \pm i\pi \int_{-\infty}^0 d\omega \rho(\omega+p{\cal E})
\end{equation}



\begin{thebibliography}{99}

\bibitem{larry} B. Gaveau and L.S. Schulman, {\it Jour. Math. Phys}
{\bf  39} (1998) 1517. \\
See also: \\
B. Gaveau and L.S. Schulman, {\it Jour. Math. Phys}
{\bf  37} (1996) 3897; {\it Phys. Lett. } {\bf A229} 347;\\
 B. Gaveau, A. Lesne and L.S. Schulman, {\it Phys. Lett. } {\bf A258}
(1999) 222.

\bibitem{Sam}  
  S. F. Edwards, in 
  {\it Granular Matter: An Interdisciplinary Approach},  
  A. Mehta ed. Springer (1994), and references therein. 
    \\
    See also: A. Mehta, R. J. Needs and S. Dattagupta; 
  {\it Journal of Statistical Physics} {\bf 68} 1131 (1992) \\
and:
    R. Monasson and O. Pouliquen,  {\it Physica A} {\bf 236} 395 (1997).



\bibitem{Beyond}
M. M{\'e}zard, G. Parisi and M. Virasoro, {\em 
Spin Glass Theory and Beyond (1987)} (Singapore: World Scientific).

\bibitem{review} 
  J-P Bouchaud, L. F. Cugliandolo, J. Kurchan and M. M{\'e}zard 
  cond-mat/9702070;
  in {\it Spin-glasses and random fields}, A. P. Young ed.
  (World Scientific, Singapore).
  
\bibitem{jorge-trieste} 
 J. Kurchan, in : {\it Jamming and Rheology: Constrained Dynamics 
  on Microscopic and Macroscopic Scales} (1997),
  http://www.itp.ucsb.edu/online/jamming2/,
  and S. F. Edwards, A. Liu and R.S. Nagel Eds., to be published\\
  J. Kurchan, cond-mat/9909306,{\it J. Phys. Condensed Matter},
to be published .
  

\bibitem{Frvi}
  S. Franz and M. A. Virasoro, 
  {\it J. Phys. A}  {\bf 33} 891 (2000).

\bibitem{KTW}
  T. R. Kirkpatrick and D. Thirumalai,
  {\it Phys. Rev. B}  {\bf 36}, 5388 (1987)
, {\it ibid} {\bf 37}, 5342 (1988),
  {\it Phys. Rev. A}  {\bf 37}, 4439 (1988). \\
  D. Thirumalai and T. R. Kirkpatrick, {\it Phys. Rev. B}  {\bf 38},
4881 (1988).
T. R. Kirkpatrick and D. Thirumalai, {\it J. Phys. A}  {\bf 22}, L149 (1989). 
  T. R. Kirkpatrick and P. Wolynes, {\it Phys. Rev. A} {\bf 35}, 3072
  (1987), {\it Phys. Rev. B}  {\bf 36},
8552 (1987). \\
  T. R. Kirkpatrick, D. Thirumalai, P. G. Wolynes, 
  {\it  Phys. Rev. A} {\bf 40}, 1045 (1989).

\bibitem{MP} 
M. M{\'e}zard and G. Parisi, {\it phys. Rev. Lett. {\bf 82} 747}  
M. M{\'e}zard and G. Parisi, 
  {\it J. Phys. Chem.} {\bf 111}  1076 (1999);
  M. M{\'e}zard, {\em First Steps in Glass Theory} {\it cond-mat/0005173}.

\bibitem{TAP} 
D.~J. Thouless, P.~W. Anderson, and R.~G. Palmer,
 {\em Phil. Mag. {\bf 35}, 593 (1977)}\\
H.~Rieger,
{\em Phys. Rev. B {\bf 46}, 14655 (1992)}\\
J.~Kurchan, G.~Parisi, and M.~A. Virasoro,
{\em J. Phys. I France {\bf 3}, 1819 (1993)}.


\bibitem{Dedominicis}
C. De Dominicis,
{\em Phys. Rep. B {\bf 67}, 37 (1980)}.


\bibitem{BM}
A. Bray and M. Moore, {\em J. Phys. {\bf C13} (1980) L469}


\bibitem{Cagipa}
A. Cavagna, I. Giardina and G. Parisi, {\em Phys. Rev. B {\bf 57}, 
11251 (1998)}.


\bibitem{crisomtap}
A.~Crisanti and H.-J. Sommers,
{\em J. Phys. I (France) {\bf 5}, 805 (1995)}.

\bibitem{Bamebu} A.~Barrat, R.~Burioni, and M.~M{\'e}zard,
{\em J. Phys. A {\bf 29}, L81 (1996)},\\
S.~Franz and G.~Parisi,
{\em J. Phys. I (France) {\bf 5}, 1401 (1995)}.

\bibitem{tapdyn}
G. Biroli, {\em J. Phys. A {\bf 32} (1999) 8365}



\bibitem{Dotsenko}
V.S. Dotsenko, M.V. Feigel'man and L.B. Ioffe, {\em Spin-glasses and 
related problems}, Soviet Scientific Reviews, vol.15 (Harwood, 1990).



\bibitem{Cuku}
L. F. Cugliandolo and  J. Kurchan,
{\it  Phys. Rev. Lett.} {\bf 71}, 173 (1993).
{\it Phil. Mag.} B {\bf 71}, 501 (1995).

\bibitem{Cukupe} 
  L.F. Cugliandolo, J.Kurchan and L.Peliti, 
  {\it Phys. Rev. E} {\bf 55}, 3898 (1997).



\bibitem{Remi} R.~Monasson,
{\em Phys. Rev. Lett. {\bf 75}, 2847 (1995)}.

\bibitem{Virasorounpu} M. A. Virasoro,
{\em unpublished}.

\bibitem{Bopari}
L.L. Bonilla, F. G. Padilla and F. Ritort, {\em Physica } {\bf A250}
(1998)
315.

\bibitem{Theo} 
  Th. M. Nieuwenhuizen, 
  {\it Phys. Rev. E} {\bf 61}  267 (2000).


\bibitem{Pa}
G. Parisi,
{\em Phys. Rev. Lett. {\bf 79}, 3660 (1997)}

\bibitem{Kobteff}
W. Kob and J.-L. Barrat,
{\em Eur. Phys. Lett. {\bf 46}, 637 (1999)}

\bibitem{aggiunta2}
M. Sellitto, {\em Eur. J. Phys. B {\bf 4}, 135 (1998)}.

\bibitem{Ru}
R. Di Leonardo, L. Angelani, G. Parisi and G. Ruocco,
 {\em cond-mat/0001311}



\bibitem{inherent}
  P.H. Stillinger and T.A. Weber 
{\it Phys. Rev. A} {\bf 25} 978 (1982).
  P.H. Stillinger and T.A. Weber,
   {\it Science} {\bf 225}, 983 (1984).    \\ 
  B. Coluzzi, G. Parisi and P. Verrocchio 
  {\it Phys. Rev. Lett.} {\bf 84} 306 (2000)\\
  F. Sciortino, W. Kob, and P. Tartaglia,
  {\it Phys. Rev. Lett.} {\bf 83} 3214 (1999),
  W. Kob, F. Sciortino, and P. Tartaglia, 
  {\it Europhys. Lett.} {\bf 49}  5906 (2000).
\\
 A. Crisanti and F. Ritort, 
{\em Activated processes and Inherent Structure dynamics 
of finite-size mean-field models for glass}, {\it cond-mat/9911226};
{\em Potential energy landscape of simple p-spin models for glasses},
{\it cond-mat/9907499}



\bibitem{enfants} 
  G. Biroli and R. Monasson, {\it Europhys. Lett. {\bf 50} (2000) 155}.



\bibitem{regis} See: R. Melin, {\it J. Physique} {\bf I 6}
  (1996) 469;   R. Melin and B. Butaud, {\it J. Physique} {\bf I 7}
  (1997) 691.


\bibitem{spectre} 
  T.W. Ruijgrok and J.A. Tjon, 
  {\it Physica} {\bf 65}  539 (1973);

  G. Biroli and R. Monasson, 
  {\it J. Phys. A}  {\bf 31} L391 (1998).




\bibitem{foot}
This follows from the fact that,
for a Fokker-Planck process:
$\langle x | e^{-t^* H } | x \rangle \sim (4 \pi T  t^*)^{-\frac{1}{2}} \;\;
 {\mbox{as}} \;\;\;  t^* \rightarrow 0$, 
a constant independent of the coordinates.


\bibitem{foot3} In the thermodynamic limit it is in fact more proper to 
compare, instead of  $t^*$,  
the timescale given by $N \left[\langle \langle H \rangle
 \rangle_{t^*}(E_o)\right]^{-1}=-\left(\frac{1}{N}\frac{\partial{\cal S}_{t^{*}}(E_{0})}{\partial t^{*}}\right)^{-1}$ (i.e. inverse average eigenvalue of the
 Fokker-Planck operator), which is the conjugate variable to $t^{*}$ under 
Legendre transformation.
  



\bibitem{ZJ}
J. Zinn-Justin, {\em Quantum Field Theory and Critical Phenomena },
 Clarendon Press 1997.



\bibitem{foot1} We should take averages over the logarithm of the
  trace.
In this kind of models, it seems safe to take instead the (annealed)
  average over the trace itself \cite{crisomtap}.



\bibitem{twogroups1}
A.J. Bray and M. Moore,
{\em Phys. Rev. Lett. (1978), {\bf 41} 1068}

\bibitem{twogroups2}
A.J. Bray and M. Moore,
{\em J. Phys. C {\bf 13} (1980), L907}


\bibitem{potters}
G. Parisi and M. Potters,
{\em Eur. Phys. Lett. {\bf 32} (1995) 13}.




\bibitem{Kurchansusy}
J. Kurchan, {\em J. Phys. France {\bf 2}, 1333 (1992)};
S. Franz and J. Kurchan, {\em Europhys. Lett. {\bf 20}, 197 (1992)}.

\bibitem{aggiunta}
A. Barrat, J. Kurchan, V. Loreto and M. Sellitto, {\em cond-mat/0006140}.

\bibitem{Ioffe}
L. B. Ioffe and D. Sherrington, {\em Phys. Rev. B {\bf 57}, 7666 (1998)}.
A. V. Lopatin and L. B. Ioffe, {\em Phys. Rev. B {\bf 60}, 6412
  (1999); 
cond-mat/9907135}.


\bibitem{Witten}
E. Witten,  {\em J. Diff. Geom} {\bf 17}, (1982) 661.


\end{thebibliography}
\end{document}